\documentclass[final,3p,twocolumn]{elsarticle}

\makeatletter
\def\ps@pprintTitle{%
  \let\@oddhead\@empty
  \let\@evenhead\@empty
  \let\@oddfoot\@empty
  \let\@evenfoot\@oddfoot
}
\makeatother

\usepackage[pagewise]{lineno}
\modulolinenumbers[1]
\usepackage{hyperref}
\usepackage{svg}
\hypersetup{colorlinks=true,pdfborder=0 0 0}
\usepackage{subcaption}
\usepackage{amsmath,bm,siunitx,amssymb,caption,booktabs}
\usepackage{ulem}

\usepackage{xspace}

\usepackage{textcomp,gensymb}
\usepackage{cleveref}
\usepackage{xcolor}

\journal{arxiv}

\begin{document}

\begin{frontmatter}

\title{Neural electron backscatter diffraction}

\address[az-mse]{Department of Materials Science and Engineering, University of Arizona, Tucson, AZ 85721, USA}
\address[az-am]{Graduate Interdisciplinary Program in Applied Mathematics, University of Arizona, Tucson, AZ 85721, USA}

\author[az-mse]{I-Tzu Huang}
\author[az-mse,az-am]{Marat I. Latypov\corref{cor1}}
\cortext[cor1]{corresponding author}
\ead{latmarat@arizona.edu}

\begin{abstract}

At the mesoscale, the state of a material is described by continuous fields. In polycrystals, crystallographic orientation and defect content vary continuously within grains, and grain boundaries trace continuous curves. Like other spatially resolved characterization methods, electron backscatter diffraction (EBSD) records this continuum on a discrete grid. Every subsequent analysis inherits the grid, whether it is classical Hough indexing or pattern-based machine learning. We introduce neural EBSD, which treats a scan as a continuous and differentiable field of Kikuchi diffraction intensity over specimen and detector coordinates. Two formulations are explored: a joint network over all four coordinates, and a factorized representation that combines continuous specimen-domain coefficient fields with learned detector-domain basis patterns. The factorized formulation exhibits higher accuracy: for recrystallized and additively manufactured Ni-base superalloys, it reconstructs \num{9e5} Kikuchi patterns per map with mean errors below \SI{1}{\percent} of the maximum intensity, while reducing data storage nearly 750-fold relative to the raw patterns. Since the learned field is continuous, patterns can be queried at any specimen position. Trained only on a quarter of the scan points, the model recovers withheld patterns whose indexed orientations fall within \SI{4}{\degree} of reference at \SI{97}{\percent} of positions in the recrystallized alloy. Analytical spatial derivatives of the differentiable representation provide a diffraction gradient that localizes grain boundaries continuously, free of indexing, disorientation thresholds, and staircase artifacts. The gradient field also exposes intragranular heterogeneity, including dislocation-cell substructure in the as-built alloy. Interactive demo: \url{https://neural-ebsd.github.io}.

% \noindent Live demo of spatially continuous reconstruction of full patterns from trained neural representation is avaialble at \url{https://neural-ebsd.github.io}

\end{abstract}

\begin{keyword} 
Electron backscatter diffraction; Kikuchi patterns; Implicit neural representation; Factorized neural representation;  Differentiable characterization; Polycrystalline materials
\end{keyword}

\end{frontmatter}

\section{Introduction}
\label{sec:intro}

At the mesoscale, the state of a material is described by continuous fields. Chemical composition, crystallographic orientation, lattice strain, and defect content -- all vary smoothly in space everywhere except interfaces. The interfaces themselves form continuous surfaces in three dimensions and curves on the observation plane. Spatially resolved characterization aims to recover these \textit{continuous} fields, yet every mapping technique, including electron backscatter diffraction (EBSD), samples them on a \textit{discrete} grid. While the discrete grid is a physical reality of experimental acquisition, we show that it does not need to be inherited into subsequent representation and analysis of fundamentally continuous microstructure fields. 

\begin{figure*}[t]
  \centering
  \includegraphics[width=\textwidth]{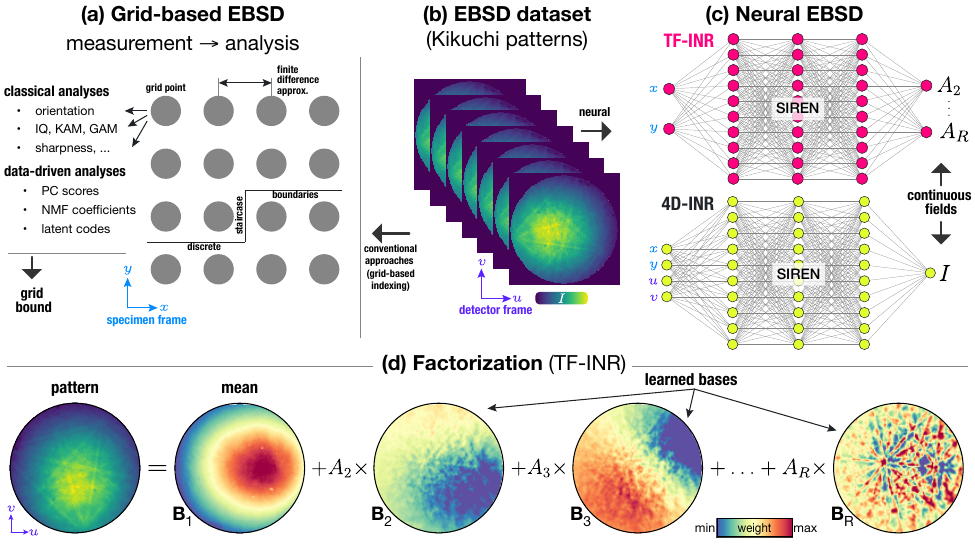}
  \caption{Visual overview of the present study: (a) grid-based EBSD analyses and some of their limitations; (b) Kikuchi patterns collected in an EBSD scan with detector frame; (c) present neural EBSD approach with SIREN networks predicting coefficient fields in TF-INR and intensity field in 4D-INR; (d) factorization as a linear combination of learned bases.}
  \label{fig:overview}
\end{figure*}

EBSD is a powerful and widely used characterization method for crystalline materials as it provides both detailed information on local lattice and statistically significant distributions over representative areas. In EBSD, a focused electron beam scans the surface of a crystalline material and produces a Kikuchi diffraction pattern (\Cref{fig:overview}b) for each scan point forming a grid \cite{schwartz2009electron}. The Kikuchi pattern contains rich information on the local structure within an interaction volume: lattice parameters, lattice distortions (e.g., from defects), crystallographic symmetry and orientation \cite{schwartz2009electron,wilkinson2012strains}. Indexing these patterns with the Hough transform \cite{kunze1993advances,krieger1996automatic} or dictionary methods \cite{chen2015dictionary} converts an EBSD scan into an orientation map. Orientation maps allow grain segmentation (by disorientation thresholding) \cite{bachmann2011grain}, calculation of geometrically necessary dislocation (GND) densities \cite{pantleon2008resolving} or kernel average misorientation (KAM) as their proxy \cite{wright2011review,kamaya2012assessment}, and grain boundary character distributions \cite{saylor2004measuring}. 

Beyond orientation, pattern-based analysis such as the image quality \cite{wright2011review}, pattern sharpness \cite{wang2023dislocation}, or cross-correlation \cite{thome2019ni} provides additional information on local lattice (im)perfection and total dislocation content. Recent data-driven work aims to preserve the entirety of Kikuchi patterns instead of reducing them to orientation or scalar metrics. Multivariate decompositions such as principal component analysis (PCA) or non-negative matrix factorization (NMF) express all patterns in a dataset as combinations of a finite number of component patterns. These approaches allow segmentation without indexing \cite{brewer2008multivariate,wilkinson2019applications,mcauliffe2020spherical,chauniyal2024employing} and accelerate dictionary indexing \cite{chen2015dictionary,varley2026accelerating}. Variational autoencoders compress each pattern into a low-dimensional latent code that reconstructs it with high fidelity \cite{calvat2025learning,liu2025learning}, while offering physical interpretation of individual latent dimensions \cite{calvat2025kikuchipattern} and indexing-free segmentation \cite{wang2026lattice}. These representations, however, are learned pattern by pattern with no reference to the position of the pattern  in the specimen frame. As a result, shuffling the scan positions would leave the learned components, latent codes, and reconstructions unchanged.

Whether classical indexing of Kikuchi patterns, advanced dislocation density estimation, or latent representation by variational encoders, all EBSD analysis workflows to date share the same fundamental limitation: they are confined to the measurement grid (\Cref{fig:overview}a). The resolution of the grid is often dictated by the practical compromise rather than the microstructure as the step size is at odds with the acquisition time and the field of view. 

This compromise introduces artifacts that propagate into microstructure analysis. For example, point-to-point disorientation thresholding \cite{bachmann2011grain} makes microstructure segmentation sensitive to the step size. A smooth intragranular gradient (e.g., in deformed or additively manufactured alloys \cite{wang2023dislocation,calvat2025learning}) may accumulate disorientation between coarsely spaced grid points sufficiently large to exceed a low-angle threshold and falsely indicate a boundary. Even for correctly detected boundaries, positions of their segments are known only to within a pixel and their traces are reconstructed as a staircase of grid-aligned segments (\Cref{fig:overview}a). The staircase approximation biases the boundary length, trace inclination, and curvature that feed grain-boundary character distributions \cite{saylor2004measuring} and chord-length statistics \cite{latypov2018application,whitman2025sr}. The dependence on step size (which again reflects measurement, not the microstructure) can similarly creep into grain size, low-angle boundary fraction, misorientation, and other microstructure descriptors. The discretization also penalizes derivative quantities as they are estimated from finite differences between neighboring pixels, which introduces noise and step-size artifacts in the resulting maps \cite{pantleon2008resolving,wright2011review,kamaya2012assessment} that persist even in high-resolution EBSD analyses \cite{jiang2013measurement}. A shared consequence in all these cases is that the grid introduces features into quantities that are then interpreted as the microstructure.

Such artifacts are not unique to EBSD and are ubiquitous in data defined on grids, including digital images consisting of pixels.  In image processing, these limitations have been tackled with implicit neural representations (INRs). An INR is a neural network trained to map coordinates to signal values so that the network itself becomes the representation of the signal replacing discrete arrays of samples \cite{sitzmann2020implicit,mildenhall2021nerf,park2019deepsdf}. Such representations have the following useful properties. Since the input is a coordinate, the signal can be evaluated anywhere, not only where it was measured. As the network is a smooth function, its spatial derivatives are available \textit{analytically} through automatic differentiation instead of finite differences. Finally, a compact set of trained network weights can replace data arrays orders of magnitude larger \cite{struempler2022implicit}. Sinusoidal representation networks (SIRENs) made these properties practical for signals with fine structure: their periodic activations capture high-frequency content and represent the derivatives of a signal with high fidelity \cite{sitzmann2020implicit}. These benefits have brought INRs from view synthesis \cite{mildenhall2021nerf} and shape modeling \cite{park2019deepsdf} into scientific and biomedical imaging \cite{xu2023nesvor,ye2023superresolution,molaei2023implicit,moussaoui2025implicit}. INRs are beginning to reach materials research as well, with recent work on locating cavities in transmission electron micrographs \cite{hsu2026inr}, parameter extraction from inelastic neutron scattering \cite{chitturi2023capturing, ni2025physics}, and accelerated four-dimensional (4D) tomography \cite{friis2025implicit}. In each case the represented signal is scalar-valued at each coordinate: an intensity in a micrograph, an attenuation in a tomographic reconstruction, or a scattering intensity in momentum–energy space. A neural representation of an entire diffraction signal nested at every position in the specimen frame has not been reported, to our knowledge. 

Extending INRs to EBSD is especially non-trivial because every scan point holds an image. The natural target for INR in this case is a 4D intensity field $I(x,y,u,v)$ over specimen coordinates $(x,y)$ and detector coordinates $(u,v)$. The two domains have very different frequency content: sharp, densely packed Kikuchi band edges in the detector frame; piecewise-smooth variation in the specimen frame, which is gentle within grains and abrupt across boundaries. The direct strategy is a single network mapping all four coordinates to intensity, which we implement as a joint formulation (4D-INR) in this study (\Cref{fig:overview}c); its shared weights, however, must handle both regimes at once. The multivariate and dictionary work above shows that the patterns in a scan can be approximated by combinations of a modest number of characteristic patterns. This observation is consistent with the results in computer vision, where factorizing high-dimensional neural fields into low-rank components has proven significantly more efficient than monolithic networks \cite{chen2022tensorf,muller2022instant}. We therefore additionally develop a tensor-factorized formulation (TF-INR), in which a SIREN models continuous coefficient fields over $(x,y)$ domain that weigh a set of learnable basis patterns held at the native detector resolution (\Cref{fig:overview}c,d). This representation is continuous where continuity enables realistic analysis and thus matters the most.

In this work, we develop these formulations and evaluate them on microstructures of a Ni-base superalloy in two states: (i) recrystallized wrought material (RX718) with equiaxed grains and sharp boundaries; (ii) as-built material from directed energy deposition (AM718) with diffuse boundaries, intragranular orientation gradients, and high defect content. The factorized formulation proves more accurate: it reconstructs the $9\times10^5$ patterns of each map with mean errors below \SI{1}{\percent} of full-scale intensity. The continuity of the learned field allows queries of patterns at positions never measured, which enables pattern-level upscaling and resolving pattern evolution along arbitrary sub-grid paths. The differentiability of TF-INR offers access to a diffraction gradient field that localizes grain boundaries continuously in the specimen frame without indexing, disorientation thresholds, and staircase artifacts, while exposing intragranular heterogeneity down to the dislocation cell substructure in the as-built alloy. 

\section{Neural EBSD}
\label{sec:framework}

\subsection{EBSD as a continuous field}

We treat a 2D EBSD scan as a discrete sampling of a continuous 4D field of the Kikuchi diffraction intensity, $I$:

\begin{equation}
I = I(x,y,u,v),
\label{eq:intensity}
\end{equation}

\noindent where ($x,y$) are coordinates in the specimen frame and ($u,v$) are the coordinates in the detector frame. Our objective is to recover this field as a continuous, differentiable function of the specimen coordinates $(x,y)$ so that the diffraction signal can be queried, differentiated, and analyzed at arbitrary locations. We address the spatial continuity with SIREN \cite{sitzmann2020implicit}, whose differentiability can be exploited for microstructure analysis (\Cref{sec:gbs}). In contrast, the detector domain can be approached with different strategies, which motivates the two formulations presented below and in \Cref{fig:overview}c.

\subsection{Two formulations of neural EBSD}
\label{sec:formulations}

Our first approach to the detector domain is to consider the detector coordinates continuous, which leads to neural approximation of the intensity field, $I(x,y,u,v)$, with a single SIREN-type network, $F_\theta$,  parametrized by weights $\theta$:

\begin{equation}
F_\theta:(x,y,u,v) \mapsto I.
\label{eq:inr}
\end{equation}

\noindent This is the most direct realization of \Cref{eq:intensity} and serves as a natural baseline, which we refer to as the \textit{joint formulation}. Given the frequency mismatch between the specimen and detector domains, a single network is not guaranteed to succeed even with the SIREN architecture. 

Since the patterns of a polycrystalline map can be well approximated by combinations of a modest number of characteristic patterns \cite{brewer2008multivariate,mcauliffe2020spherical,chen2015dictionary,varley2026accelerating}, we alternatively decompose the diffraction signal as

\begin{equation}
I(x,y,u,v)=\sum_{r=1}^{R} A_r(x,y)\,B_r(u,v),
\label{eq:factorized}
\end{equation}

\noindent where $A_r(x,y)$ is the $r^\text{th}$ coefficient field in the specimen domain, $B_r(u,v)$ is the $r^\text{th}$ basis pattern in the detector domain, and $R$ is the rank of the decomposition. The bases carry the characteristic diffraction patterns whose weighted combination represents the diffraction signal across the map. The coefficients $A_r(x,y)$ serve as the weights of the corresponding bases at every scan position (see \Cref{fig:overview}d). 

With the specimen--detector decomposition, the neural representation can be pursued with a neural network that predicts coefficient fields, $\mathbf{A}$: 

\begin{equation}
f_{\theta}: (x,y) \mapsto (A_1,A_2,\ldots, A_R), 
\label{eq:separate}
\end{equation}

\noindent optimized jointly with the corresponding basis patterns, $\mathbf{B}$. The basis patterns are parameterized as $R$ learnable weight arrays of size $H_p \times W_p$ matching the size and resolution of the EBSD detector screen producing patterns of height $H_p$ and width $W_p$ (in pixels). The detector domain is therefore represented at its native acquisition resolution, while the specimen domain is represented continuously through the learned coefficient fields $\mathbf{A}(x,y)$. When restricted to the acquisition grid, \Cref{eq:factorized} reduces to the rank-$R$ decomposition that underlies PCA and NMF analyses of Kikuchi patterns \cite{brewer2008multivariate,mcauliffe2020spherical}. The difference is that the $\mathbf{A}$ coefficients are the output of a continuous function of specimen position instead of being free per-point parameters. This distinction turns a dimensionality reduction scheme into a field that can be queried and differentiated anywhere within the measurement domain. Both formulations satisfy our objective of spatial continuity in the specimen domain. Practical architectural choices with the dataset-specific implementation details are given in \Cref{sec:methods}.

\subsection{Training strategy}

Both formulations are trained by minimizing a reconstruction loss over sampled specimen--detector coordinates with a few additional considerations specific to EBSD. First, since an EBSD dataset can contain hundreds of millions of intensity values across the combined 4D domain, we sample intensities from a subset of 4D coordinates, $\Omega$, at each training iteration. Second, Kikuchi patterns contain high-frequency, narrow band edges with a slowly varying background \cite{britton2016tutorial}. As band edges carry the most important information, the corresponding pixels should be emphasized during learning so that the network does not focus on easier to predict yet less informative background regions of the Kikuchi diffraction. For these reasons, we identify band regions as part of pattern pre-processing (see Supplementary Material) and then, during training: (i) introduce weights, $w(u,v)\ge0$, that emphasize the intensity at band-region pixels and (ii) bias the sampling to include more band-region pixels in the $\Omega$ subsets. These two measures are complementary as the sampling bias focuses training to informative regions, while pixel weighting ensures that the gradient signal from band pixels is not diluted by background pixels within a given iteration.

\begin{figure*}[t]  % FINALIZED ! 
  \centering
  \includegraphics[width=\textwidth]{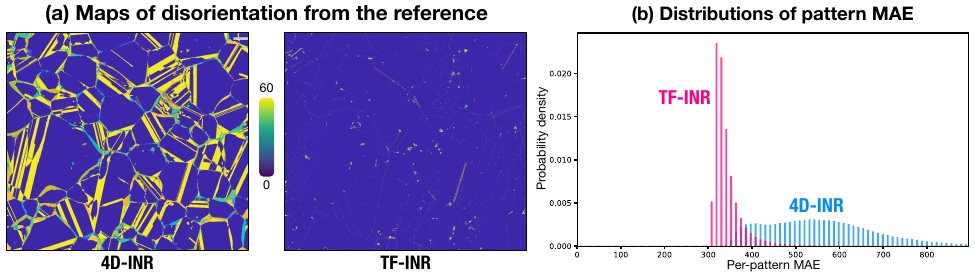}
  \caption{Reconstructions of the RX718 map with 4D-INR and TF-INR compared in terms of (a) orientation deviation maps and (b) distributions of pattern-level MAE values.}
  \label{fig:inr}
\end{figure*}

With these considerations, the reconstruction loss is defined as

% \begin{equation}
% \mathcal{L}_\text{pixel} = \frac{1}{|\Omega|} \sum_{(x,y,u,v) \in \Omega} w(u,v) \times \left( \hat{I}(x,y,u,v) - I(x,y,u,v) \right)^2,
% \end{equation}

\begin{multline}
\mathcal{L}_\text{pixel} = \frac{1}{|\Omega|} \sum_{(x,y,u,v) \in \Omega} w(u,v) \\
\times \left( \hat{I}(x,y,u,v) - I(x,y,u,v) \right)^2,
\end{multline}

\noindent where the predicted intensity, $\hat{I}$, depends on the formulation:

\begin{equation}
\hat{I}= 
\begin{cases} 
F_\theta(x,y,u,v) & \text{(4D-INR)}, \\ \sum_{r=1}^{R} [f_\theta(x,y)]_r \, B_r(u,v) & \text{(TF-INR)}.
\end{cases}
\label{eq:pred_intensity}
\end{equation}

The joint formulation minimizes the reconstruction loss as the total training objective: $\mathcal{L}_\text{total} = \mathcal{L}_\text{pixel}$. 

For the factorized formulation, we introduce an additional term, $\lVert \mathbf{B} \rVert_F^2$, that regularizes the basis patterns to prevent the ambiguity between $(A_r,B_r)$ and $(cA_r,B_r/c)$ for any $c\ne0$ in \Cref{eq:factorized}:

\begin{equation}
\mathcal{L}_\text{total} = \mathcal{L}_\text{pixel} + \lambda_B \, \|\mathbf{B}\|_F^2,
\label{eq:loss_total}
\end{equation}

\noindent The term $\lambda_B \, \|\mathbf{B}\|_F^2$ penalizes large basis-pattern magnitudes and thus prevents the optimizer from inflating $B_r$ while shrinking $A_r$ during training ($\|\cdot\|_F$ is the Frobenius norm). 

Training is performed for each individual EBSD dataset by adjusting: (i) SIREN weights $\theta$ in the joint formulation; (ii) both SIREN weights $\theta$ and basis-pattern weights $\mathbf{B}$ in the factorized formulation.

\section{Materials and methods}
\label{sec:methods}

\subsection{Materials and EBSD data}

We demonstrate the framework on EBSD datasets from polycrystalline Ni-base superalloys obtained and published by Calvat et al.\ \cite{calvat2025kikuchipattern}. Specifically, Inconel 718 (IN718) alloys in two distinct microstructure states were considered here (RX and AM) to sample microstructural diversity that should be handled by the neural EBSD framework. RX718 is a recrystallized wrought IN718 alloy with well-defined grains and sharp boundaries; AM718 is an additively manufactured IN718 alloy in as-built condition (after directed energy deposition) with less distinct boundaries, intragranular heterogeneity, and the elevated defect content. Each map covers \SI{0.9}{\milli\meter\squared} surface area at a \SI{1}{\micro\meter} step (\num{9e5} patterns), which allows testing the scalability of the neural representation beyond sizes of typical EBSD maps. We train on the $120\times120$ patterns as 16-bit images provided with the dataset, which preserve the characteristic diffraction features of original $480\times480$ images at reduced computational cost. 

The datasets were collected \cite{calvat2025kikuchipattern} on a Thermo Fisher SEM  equipped with an EDAX OIM-Hikari detector with an accelerating voltage of \SI{20}{\kilo\volt}, current of \SI{6.4}{\nano\ampere}, and exposure time of \SI{8.5}{\milli\second} per diffraction pattern at \SI{12}{\milli\meter} of working distance and a \SI{70}{\degree} tilt. The full  details on the data acquisition as well as the chemical composition and processing history of the alloys can be found in Calvat et al.\ \cite{calvat2025learning,calvat2025kikuchipattern}.

\subsection{Architecture and training}

The two formulations in \Cref{sec:formulations} used distinct architectures but shared some of the training settings.

\paragraph{4D-INR} The 4D-INR network implementing the joint formulation (\Cref{eq:inr}) had five SIREN layers of width $256$: input layer with frequency $\omega_0^{(1)}=30$, and hidden layers with frequency $\omega_0^{(h)}=30$. To increase the capacity of the monolithic network for band structure in the detector domain, the detector coordinates $(u,v)$ were additionally lifted using a random Fourier feature map with $m=256$ frequencies sampled from $\mathcal{N}(0,\sigma^2)$ with $\sigma=3.0$. The Fourier features were concatenated with the raw coordinates, which results in an effective input dimensionality of $4+2m=516$.

To stabilize optimization of the joint 4D field, training used a scaling curriculum for detector coordinates. Specifically, the detector coordinates were rescaled by a factor that increases linearly from $0.25$ to $1.0$ over the first \num{2e4} optimization steps so that the model first fits the coarse pattern structure before resolving the full detector coordinate variation.

\paragraph{TF-INR} In TF-INR implementing the factorized representation (\Cref{eq:factorized}), the coefficient field $\mathbf{A}(x,y)$ was parameterized by five SIREN layers of width $512$, with the first layer frequency $\omega_0^{(1)}=30$, and hidden layer frequency $\omega_0^{(h)}=30$. The network outputs $R-1$ learned coefficients with leading coefficient kept constant and equal to unity, $A_1=1$, everywhere in the specimen domain. The rank was fixed at $R=512$ in this study.

The detector basis was represented explicitly as a learnable tensor $\mathbf{B}\in\mathbb{R}^{R\times H_p\times W_p}$ ($H_p=W_p=120$ in pixels). %, where $H_p=120$ and $W_p=120$ are the height and width of the patterns used in this study. 
The first basis, $\mathbf{B}_1$, is initialized as the empirical mean Kikuchi pattern, so $A_1=1$ preserves the average pattern across the dataset. The remaining bases were initialized from leading principal components obtained with PCA of a random subsample of up to \num{2e5} training patterns. All basis images were then optimized jointly with the SIREN weights. This strategy thus used a scan-specific initialization with the dominant Kikuchi pattern modes while keeping bases fully trainable. TF-INRs were trained following the objective in \Cref{eq:loss_total} with $\lambda_B=10^{-6}$.

Both 4D-INR and TF-INR formulations used Adam optimization with a learning rate of $10^{-4}$ and a batch of \num{32768} sampled specimen--detector coordinates. The models were trained up to \num{2e5} steps with early stopping after 30 consecutive evaluations (every \num{1000} steps) without an improvement greater than $\delta=10^{-5}$. Detector pixels identified as band regions during pre-processing were assigned weight $w=2$ in $\mathcal{L}_\text{pixel}$. All remaining pixels were assigned $w=1$. 

\subsection{Two-level evaluation}

The fidelity of the trained neural representations against the original measurements was evaluated at two levels: pattern and indexed orientation. At the pattern level, we used two complementary metrics: mean absolute error (MAE) to capture pixel-level intensity accuracy and structural similarity index measure (SSIM) to characterize the pattern as a whole. MAE captures the absolute difference between the reconstructed and measured Kikuchi patterns averaged over the detector pixels within the aperture mask (\Cref{eq:mae_pattern}, \Cref{fig:methods}b). MAE compares detector pixels independently and thus overlooks the spatial arrangement of the Kikuchi bands. SSIM complements it by comparing local means, variances, and covariances \cite{wang2004image}. Evaluated over local windows within the aperture mask (\Cref{eq:ssim}), SSIM captures faithfulness of the band configuration and contrast -- key pattern features that carry most physical information. %These two metrics are thus complementary as MAE captures pixel-level intensity accuracy, whereas SSIM characterizes the pattern as a whole.

At the orientation level, we indexed both the reconstructed and measured patterns with conventional Hough-transform algorithms (implemented in \texttt{kikuchipy} \cite{anes2026kikuchipy, wright1992automatic}) and computed the symmetry-reduced disorientation angle between the two orientations (using \texttt{MTEX} \cite{bachmann2010texture, krakow2017three}). We refer to this angle as the \textit{orientation deviation} (\Cref{eq:disorientation}). It quantifies the impact of reconstruction errors on the indexed orientation as the most common output of EBSD analysis. The orientation deviation isolates errors of crystallographic significance from those affecting only absolute detector-pixel intensities. 

\begin{figure*}[h] % FINALIZED ! 
  \centering
  \includegraphics[width=\textwidth]{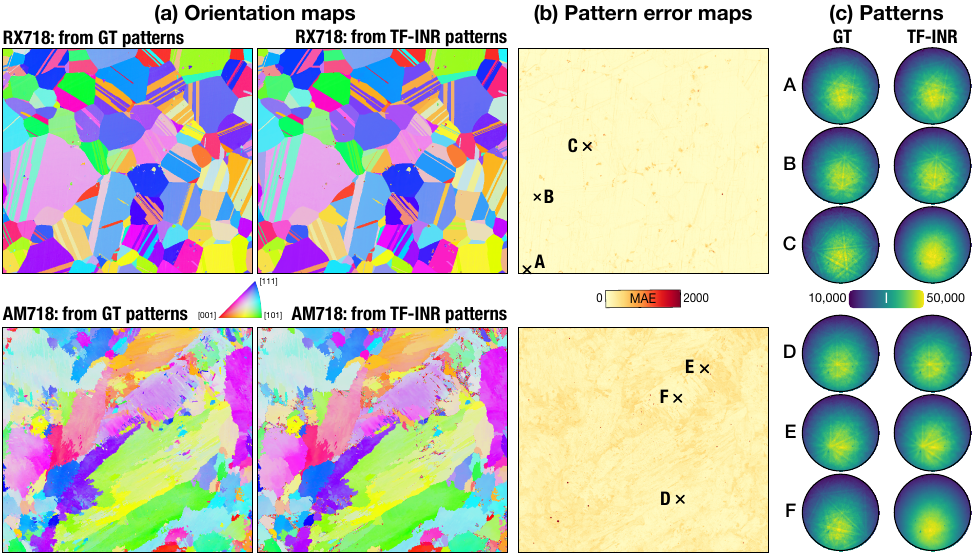}
  \caption{Reconstruction of EBSD data evaluated in terms of (a) orientation maps obtained with Hough indexing of measured (GT) Kikuchi patterns and those reconstructed from TF-INR and (b) error maps showing pattern-level MAE (\Cref{eq:mae_pattern}). Visual comparison of (c) measured and reconstructed patterns for selected grid points are also shown corresponding to locations with both low and high pattern-level MAEs.} 
  \label{fig:recon}
\end{figure*}

As all error metrics are computed for each grid point, ($x,y$), the reconstruction accuracy for an EBSD dataset is reported through statistical summaries, distributions, and maps.  Statistical summaries include map-averaged values and fractions of the grid points with an error metric below (or above) a threshold: e.g., fraction of grid points with the orientation deviation below \SI{4}{\degree} based on the definition of \textit{small} disorientation introduced elsewhere  \cite{ding2020indexing}. Similarly, the map-averaged MAE is the mean of pattern MAE over all scan positions (\Cref{eq:mae_map}). MAE is reported in 16-bit intensity counts; dividing by the dataset-wide maximum intensity (\num{61183} for RX718, \num{58727} for AM718) gives the error as a percentage of full scale. 

\section{Results: Neural EBSD of superalloys} 
\label{sec:results}

\subsection{Reconstruction}
\label{sec:recon}

\textit{4D-INR vs TF-INR}. We first benchmark the joint and factorized formulations on the RX718 map, where the well-defined recrystallized grain structure can clearly test the reconstruction fidelity. The two formulations produce significantly different reconstructions (\Cref{fig:inr}). The 4D-INR baseline yields a broad MAE distribution shifted toward higher reconstruction errors (\Cref{fig:inr}b), and its orientation deviation map shows angular errors extending across grain interiors. A large fraction of grid points, \SI{19.1}{\percent}, exceeds the \SI{4}{\degree} threshold. TF-INR, in contrast, produces a narrow MAE distribution in a lower-value range and a uniform orientation deviation map with errors concentrated at grain boundaries. The fraction of grid points with orientation deviation \qty{\ge4}{\degree} is only \SI{1.2}{\percent} (of \num{9e5} orientations). Since TF-INR offers a much more faithful reconstruction, we focus on this formulation for the rest of this study.

\begin{figure*}[t] % FINALIZED !  
  \centering
  \includegraphics[width=\textwidth]{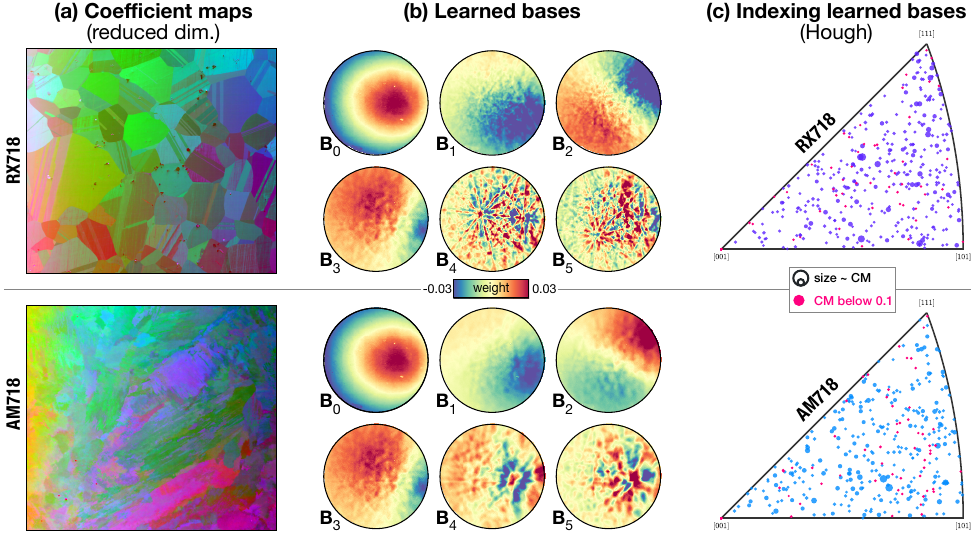}
  \caption{Learned TF-INR representation: (a) coefficient field $\mathbf{A}(x,y)$ projected from $R=512$ to three dimensions with PCA shown with RGB; (b) examples of learned bases, $\mathbf{B}_r$; (c) 512 crystallographic orientations obtained by Hough indexing of the learned bases (circle size is proportional to indexing confidence metric, orientations obtained with confidence metric below 0.1 are shown in red).} 
  \label{fig:bases}
\end{figure*}

The reconstruction fidelity of TF-INR is next examined for both microstructures, RX718 and AM718, in terms of indexed orientation maps, MAE maps, and visual inspection of selected individual patterns (\Cref{fig:recon}) as well as SSIM analysis. The orientation maps (from reconstructed patterns) closely match the reference orientation maps (from measured patterns) for both microstructures. The map-averaged MAE values are 344 and 483 intensity counts for RX718 and  AM718, which correspond to normalized MAE of \SI{0.6}{\percent} and \SI{0.8}{\percent} relative to the dataset-wide maxima. The spatial distribution of the residuals mirrors the microstructural complexity of each sample. In RX718, the larger errors in both intensity and indexed orientation are confined to small grains (or particles) and grain boundaries. In AM718, the residuals are more uniformly distributed and reach into the grain interiors, consistent with its stronger intragranular heterogeneity. The representative patterns in \Cref{fig:recon}c show examples of grid points with low and high MAE. The low-MAE patterns (A, B, D, E) display excellent agreement in both the Kikuchi-band configuration and intensity. At the high-MAE points (C, F), the TF-INR model predicts mostly background intensity with weak bands, whereas the measured patterns have clear bands. SSIM analysis confirms that TF-INR reconstruction is of high fidelity  in terms of not only mean intensity but also band configuration and contrast: map-averaged SSIM values are 0.97 for RX718 and 0.94 for AM718. SSIM above 0.9 is obtained for all \num{9e5} patterns of RX718 (with a sharp peak at 0.97) and for most patterns of AM718 (see \Cref{fig:ssim} in Supplementary). 

Given strong reconstruction performance of TF-INR on both maps, we inspect the learned representation by visualizing the coefficient fields and selected basis patterns (\Cref{fig:bases}).  The coefficient field $\mathbf{A}(x, y)$ contains a 512-dimensional vector at every grid point. To visualize the field, we project the coefficients to three dimensions by PCA and map the result to RGB (\Cref{fig:bases}a). The projection shows spatially coherent regions corresponding to the grain structure in both alloys (compare with \Cref{fig:recon}a). It means the continuous spatial field has captured the piecewise-smooth organization of the polycrystalline microstructure without being explicitly given any orientation or boundary information. Many of the learned bases $\mathbf{B}_r$ resemble Kikuchi patterns instead of being abstract (\Cref{fig:bases}b).  The leading basis, $\mathbf{B}_1$, is smooth and similar to background intensity distribution because it is initialized from the mean pattern over the dataset (its coefficient, $A_1$, is fixed to unity). Other bases include finer band structures that appear as realistic diffraction patterns bearing crystallographic orientations. Indeed, Hough-indexing of the bases (like regular Kikuchi patterns) results in valid orientations with confidence metric (see Supplementary) above 0.1 for many of the 512 bases (\Cref{fig:bases}c).

\subsection{Continuous query of the diffraction field}
\label{sec:interp}

A unique advantage of the neural representation is that Kikuchi patterns can be reconstructed at arbitrary specimen coordinates not restricted to the original measurement grid. Here, we provide two demonstrations of this property, including (i) coarse-grid super-resolution: reconstruction of Kikuchi patterns on a fine grid from sparse sampling, and (ii) sub-grid path sampling: pattern evolution along arbitrary paths with sub-grid resolution.

\begin{figure*}[h!]
  \centering
  \includegraphics[width=\textwidth]{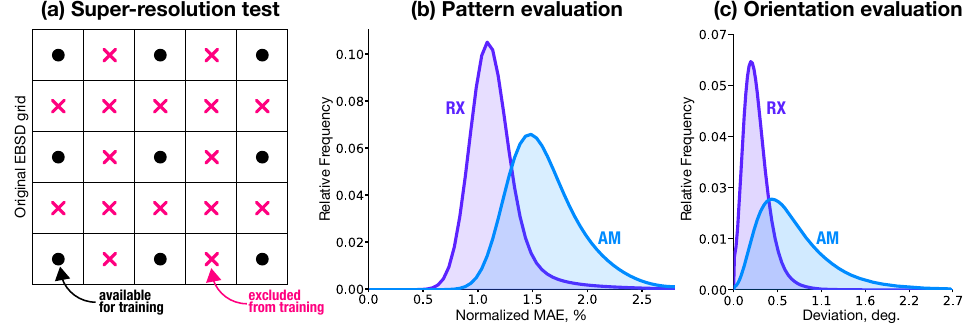}
  \caption{Demonstration of super-resolution: (a) hold-out of every other pixel (red cross) during training to emulate coarse-grid experimental measurement (black circles); (b) distributions of normalized MAE values for \num{6.75e5} held-out grid points; (c) distributions of orientation deviations from reference orientations at held-out grid points.}
  \label{fig:super}
\end{figure*}

\paragraph{Coarse-grid super-resolution} We first test the capability of the TF-INR model to upscale coarse-grid measurement to a high-resolution grid. To this end, we re-trained the TF-INR model using only every other scan position in both horizontal and vertical directions of each map. For $900\times1000$ maps, using only a quarter of points reduces the number of scan positions available for training to \num{225000}. The remaining \num{675000} positions were unseen by the model and used only for evaluation. After training on this sparse grid, TF-INR was queried at the withheld positions to reconstruct the missing patterns and evaluated in terms of orientation deviation with the same procedure as for the full-grid reconstruction (\Cref{sec:recon}). 

Despite being trained on only \SI{25}{\percent} of the original scan positions, TF-INR successfully recovers the withheld patterns with high accuracy in terms of indexed orientation. For RX718, \SI{97}{\percent} of the held-out points were reconstructed within $4^\circ$ of the reference orientation, with a mean orientation deviation of \SI{0.9}{\degree}. For AM718, the corresponding fraction was \SI{89}{\percent}, with a mean orientation deviation of \SI{4.7}{\degree}. The lower accuracy in AM718 is consistent with its more heterogeneous microstructure, stronger intragranular orientation variation, and larger population of mixed patterns associated with boundaries or defect substructure. Nevertheless, the result shows the potential of TF-INR to upscale EBSD maps at the Kikuchi pattern level from coarse measurements with only \SI{3}{\percent} of fine-grid points having significant deviations in terms of the crystallographic orientation (as demonstrated by the RX718 case). 

\begin{figure*}[h!]  % FINALIZED ! 
  \centering
  \includegraphics[width=\textwidth]{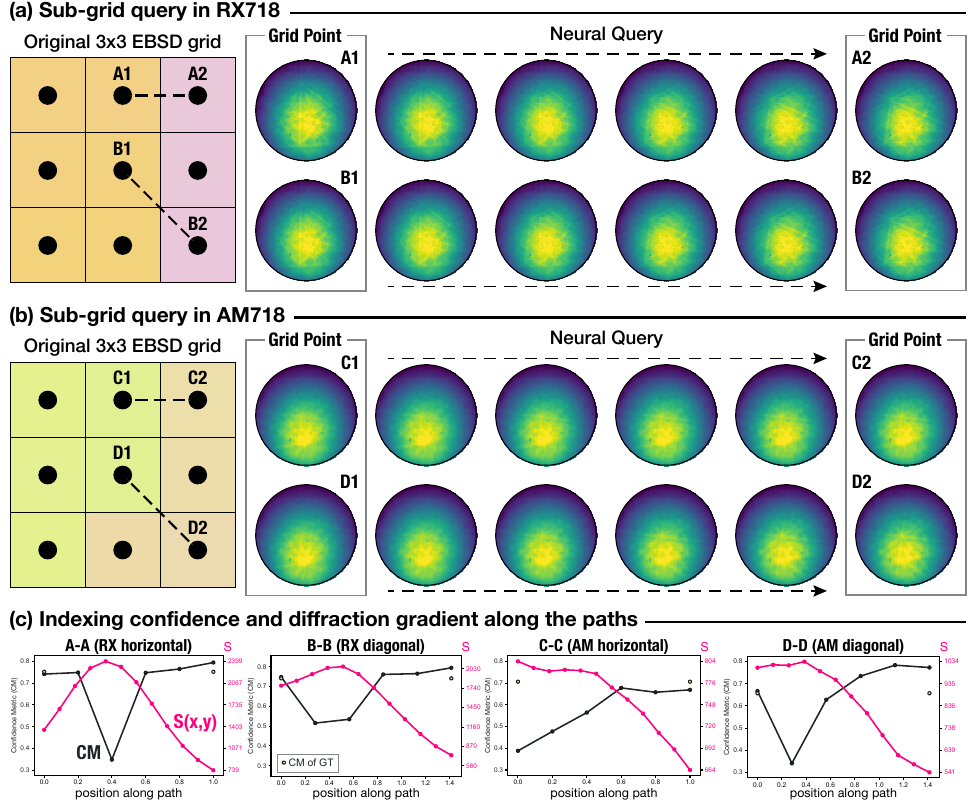}
  \caption{Demonstration of sub-grid path sampling: (a-b) sub-grid query between EBSD pixels in two directions in RX718 and AM718 microstructures; (c) indexing confidence and diffraction gradient (\Cref{sec:gbs}) along the paths. See \Cref{fig:methods}c for the locations of the $3\times3$ grids shown here.}
  \label{fig:interp}
\end{figure*}

\paragraph{Sub-grid path sampling} To probe TF-INR at resolutions finer than the original measurement step, we queried the learned field at four equally spaced positions between adjacent grid points. This sub-grid sampling corresponds to five-fold refinement in spatial resolution along a chosen path. Each pair was selected to span two grains to analyze how the model handles sharp orientation changes at sub-grid scale. This is the regime most relevant to grain boundary characterization, where conventional grid-based analysis is intrinsically limited by the measurement resolution. Two path orientations were considered for each microstructure: horizontal (aligned with one of the scan axes) and diagonal (not aligned with the scan axes).

Along each path, the reconstructed patterns transition smoothly from one endpoint to the other (\Cref{fig:interp}a,b). Diffraction features associated with the first grain progressively weaken while features of the second grain emerge. This behavior is consistent with physical diffraction across a grain boundary \cite{shi2021indexation,tong2015effect}. The transition is non-uniform along the path, which is seen from visual inspection of the patterns and with the indexing confidence metric (\Cref{fig:interp}c). The confidence metric is low when the Hough peak structure is ambiguous, including the overlap of patterns due to contributions of two (or more) grains at a grain boundary \cite{tong2015effect, marquardt2017quantitative}. Tracked for the queried patterns, the confidence metric exhibits a dip along three of the four paths. In each case, the dip is off the path midpoint. Since confidence drops where patterns mix contributions from neighboring grains, the minimum in the curve likely marks a boundary at sub-grid resolution. The diagonal path in the RX718 case (B-B in \Cref{fig:interp}a,c) shows that the low-confidence region is not necessarily confined to a single sub-grid point.  These results suggest the potential of localizing grain boundaries in a continuous fashion in the specimen frame (pursued in the next section). 

The remaining path in AM718 (C-C in \Cref{fig:interp}b,c) exhibits a different behavior. The TF-INR reconstruction at one endpoint of this path has a relatively high error (MAE = 432 counts) compared to the rest of the map and produces a low-confidence pattern despite high confidence in indexing the measured pattern at the same position (see circle in \Cref{fig:interp}c). Along the path, the model recovers patterns of progressively higher confidence towards the other endpoint (\Cref{fig:interp}b). This result demonstrates smooth behavior of the continuous representation even in a scenario of locally imperfect reconstruction.

To test whether these sub-grid patterns could be obtained without neural EBSD, we generated patterns along the same paths by bilinear and bicubic interpolation between the measured endpoint patterns and indexed all three sequences (\Cref{fig:supp_interp-comp} in Supplementary). Although the interpolated patterns are visually similar to the TF-INR reconstructions (\Cref{fig:supp_interp-comp}a), their confidence profiles are clearly different (\Cref{fig:supp_interp-comp}b). Both interpolations reach their minima at the midpoint and lose confidence symmetrically about it. By contrast, the TF-INR minima fall off-midpoint (around 0.42 for RX718 and 0.38 for AM718), at the same positions at which the diffraction gradient peaks (\Cref{fig:interp}c). The symmetry of the interpolated profiles is a property of the operation that does not necessarily reflect the microstructure. Indeed, a pixel-wise blend of two endpoint patterns is mixed most strongly halfway between them, so its confidence minimum reports the geometry of the sampling and carries no information about the location of the actual boundary. The TF-INR profile is under no such constraint, and its minimum is set by the learned field rather than by the prescribed mixing of the two endpoint patterns.

\subsection{Continuous and indexing-free grain boundary localization}
\label{sec:gbs}

\begin{figure*}[h]  % FINALIZED ! 
  \centering
  \includegraphics[width=\textwidth]{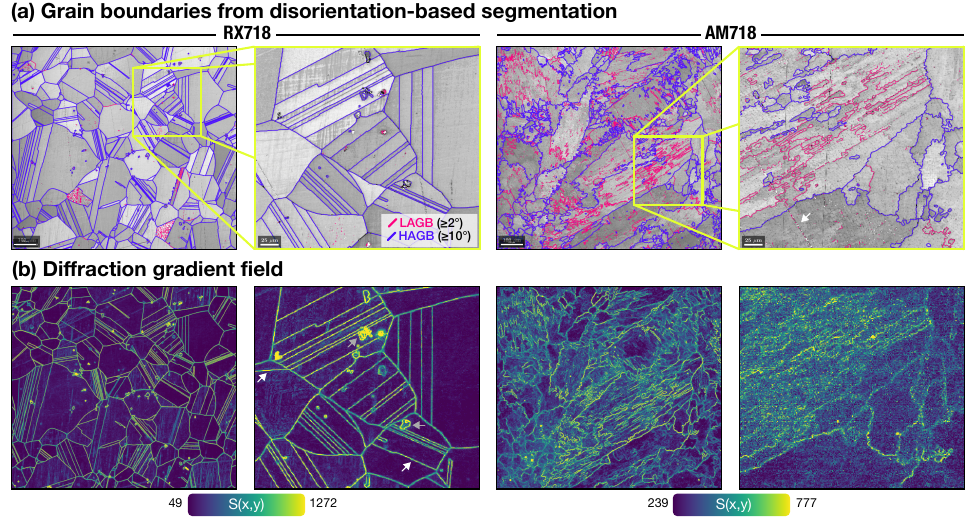}
  \caption{Grain boundary analysis for RX718 and AM718 microstructures: (a) conventional disorientation-based segmentation and (b) diffraction gradient field from differentiation of the coefficient field (\Cref{eq:sensitivity}). Grain boundaries in (a) are shown for two thresholds: \SI{2}{\degree} (low angle, LAGB) and \SI{10}{\degree} (high angle, HAGB) overlaid with the image quality map from experimental EBSD measurement. Arrows indicate interesting features, including (i) white arrows in (b): examples of diffraction gradient varying along a single boundary; (ii) gray arrows in (b): small areas enclosed by high diffraction gradient possibly related to particles; (iii) white arrow in (a): polishing artifact detected as a LAGB with disorientation thresholding that is absent in the diffraction gradient field in (b). }
  \label{fig:gbs}
\end{figure*}

The differentiability of the neural EBSD representation enables a new approach to grain boundary localization and visualization. Instead of pixel-wise thresholding of disorientations on a discrete grid, boundaries can be localized directly from analytical spatial derivatives in the specimen domain. For TF-INR specifically, we introduce the diffraction gradient field expressed as

\begin{equation}
    % S(x,y) \;=\; \sqrt{\, \sum_{r=1}^{R} \left\|\, \nabla_{\!x,y}\, A_r(x,y) \,\right\|^2 \,},
    S(x,y) = \sqrt{\, \sum_{r=1}^{R} \left [  \left ( \frac{\partial A_r}{\partial x}\right )^2  + \left(\frac{\partial A_r}{\partial y}\right)^2 \right] } = \left\| \nabla_{\!x,y} \mathbf{A} \right\|_F,
    \label{eq:sensitivity}
\end{equation}

\noindent which quantifies the magnitude of spatial variation of the coefficient field, $\mathbf{A}(x,y)$. Since $\mathbf{A}(x,y)$ are represented by a SIREN, the gradients $\nabla_{\!x,y}\mathbf{A} $ are accessible analytically by automatic differentiation without the need for finite-difference approximation, which is susceptible to noise that plagues gradient-based analyses on conventional EBSD grids. 

From \Cref{eq:sensitivity}, the magnitude of the gradient is high when there is a significant change in the coefficients $\textbf{A}$, which reflects sharp discontinuity of the diffraction field in the specimen domain. Such discontinuities are expected at grain (or other) boundaries and locations with changes in the lattice defect content. The field, $S(x,y)$, thus offers a new route for indexing-free localization of grain boundaries and other defects that can be visualized with any resolution. 

Visualized for RX718 and AM718 microstructures, the diffraction gradient maps exhibit boundary networks largely consistent with the grain boundaries obtained based on disorientations (\Cref{fig:gbs}). While most conventional workflows for grain segmentation and boundary visualization rely on discrete thresholds (e.g., \SI{2}{\degree}, \SI{10}{\degree} as shown in \Cref{fig:gbs}a), the diffraction gradient is intrinsically continuous and does not require selection of a threshold. 

The continuous diffraction gradient reveals that the discontinuity magnitude is non-uniform even along a single grain boundary with variations along its trace (\Cref{fig:gbs}). Most segmentation workflows used in practice only detect the presence of a boundary for a chosen threshold and discard this variation, even though the underlying disorientation is continuous. The diffraction gradient map reveals these variations explicitly with a single scalar without the need for advanced coloring schemes (e.g., \cite{patala2011continuous}). Since $S$ captures the full change in the diffraction patterns not limited to the orientation (e.g., sharpness), the variation along the boundary may reflect subtle changes in the grain boundary character (e.g., inclination) or its defect content. The diffraction gradient also localizes boundaries of particles: some with clean interiors, some filled with high $S$ (\Cref{fig:gbs}b, gray arrows). The high-$S$ regions could correspond to particles with rough surfaces and deteriorated or significantly varying patterns within their areas. 

The continuously varying magnitude also reveals intragranular heterogeneities. In the RX718 case, the diffraction gradient field exhibits cross-hatched patterns of slightly elevated magnitude. Based on the corresponding image quality maps, these features are associated with lattice defects introduced with remnant polishing scratches. In contrast, polishing scratches (some of which are conventionally segmented as low-angle grain boundaries, see arrow in the AM718 map in \Cref{fig:gbs}a) do not appear in the diffraction gradient maps for AM718. The exposure of intragranular heterogeneity is most evident in AM718. For this microstructure, the diffraction gradient map reveals high-angle grain boundaries, consistent with the conventional thresholding, a network of low-angle grain boundaries, as well as elevated gradient magnitudes throughout the grain (and subgrain) interiors. The elevated magnitudes most likely originate from dislocation cells reported for this alloy in the as-built condition \cite{calvat2025learning}. 

The gradient field is also continuous in the specimen domain and can be sampled with automatic differentiation  on grids of any resolution. As alluded in the previous section, this property allows localizing grain boundaries in a spatially continuous fashion beyond grid midpoint positions or staircase-like approximations characteristic of grid-based segmentation. The boundary continuity is illustrated in the diffraction gradient profiles for the same paths analyzed in \Cref{fig:interp}. These profiles show peaks in the same off-midpoint positions as dips in the confidence metric (\Cref{fig:interp}c). 

\section{Discussion} 
\label{sec:disco}

 Neural EBSD changes the object of analysis: instead of a discrete set of independent patterns attached to a grid, an EBSD scan becomes a single diffraction field, continuous and differentiable. Each capability demonstrated above is a natural property of this field that does not require a separate algorithm. Reconstruction is evaluation of the field on the acquisition grid, super-resolution is evaluation off the grid, boundary localization is differentiation. This section discusses what the neural representation reveals about the microstructure; what it implies for acquisition, storage, and sharing of EBSD data; its present limitations and outlook. %how the construction transfers beyond EBSD (§5.3), 
 
 \subsection{New capabilities for microstructure analysis}
 \label{sec:disco-1}

The diffraction gradient $S(x, y)$ most clearly illustrates the advantages of differentiability. It locates boundaries by identifying the locations of fastest spatial changes in the diffraction field without the need in indexing or disorientation thresholds. The diffraction gradient also needs no grid as the derivatives of the coefficient field are available analytically at any specimen position. The sub-grid paths (\Cref{fig:interp}) demonstrate sub-pixel localization of boundaries achieved with this grid-free approach. Peaks in $S$ coincide with dips in the indexing confidence metric at the same off-midpoint positions, so two independent probes place the boundary at the same sub-grid location. % SUPP
This result confirms that a boundary position has continuous coordinates instead of an assignment to the gap between two pixels. Likewise, boundary traces are not limited to staircases of grid-aligned segments. While the demonstrations here are visual, converting them into quantitative analysis (of grain size distributions, boundary length per unit area, trace inclinations for boundary-character statistics) can be readily achieved by segmenting $S$ with watershed-type algorithms \cite{beucher1992watershed,fotos2023deep}. A profile of $S$ in \Cref{fig:supp_gradient} suggests segmentation feasibility. % SUPP

Since $S(x,y)$ is continuous along a boundary as well as across it, it preserves information discarded with thresholded segmentation. The discontinuity magnitude varies visibly along individual boundary traces (\Cref{fig:gbs}b, white arrows). As $S$ aggregates every source of pattern change (orientation jump, band sharpness, diffuse contrast), this variation likely encodes changes in boundary character, local inclination, or near-boundary defect state. %We offer it as a hypothesis that the representation makes testable, not as a settled interpretation.
The analysis of the as-built microstructure shows the value of a scalar that reflects more than orientation changes. In AM718, the gradient field recovers the high-angle boundary network similar to conventional segmentation, adds the low-angle subgrain network, and reveals elevated magnitude throughout grain interiors (\Cref{fig:gbs}). The intragranular signal is consistent with the dislocation-cell substructure documented for this material in the as-built condition \cite{calvat2025learning}. Such substructure contributes to strength through cell-wall hardening and evolves strongly under heat treatment in additively manufactured alloys \cite{wang2018additively}. A single field that exposes cells, subgrain walls, and grain boundaries from raw patterns in a single autodifferentiation pass can serve as a microstructural fingerprint of the build with potential use in part qualification and tracking heat-treatment response \cite{seifi2016overview,arciniaga2026insights}.

Encoding the whole field can be also robust to local corruption of the signal. In AM718, polishing scratches that disorientation thresholding misclassifies as low-angle boundaries (\Cref{fig:gbs}a, arrow) leave no trace in $S$. In RX718 however, the same type of artifact does appear as faint cross-hatching of elevated $S$ within grains. The asymmetry has a simple interpretation based on the shared rank $R$ used in this study. With $R=512$, the captured variance is near the \SI{95}{\percent} target for RX718 but lies below it for AM718 (\Cref{fig:methods}a). So the TF-INR model for RX718 has a capacity to capture artifacts, whereas the model for AM718 leaves them out. Rank, therefore, can act as an implicit noise filter -- a behavior with practical uses and limitations further discussed below.

\subsection{Implications for EBSD data acquisition and management}
\label{sec:disco-2}

The continuity of the field can enable efficient strategies of experimental acquisition. Trained on one quarter of the scan positions, TF-INR successfully recovers withheld patterns on a finer grid (\Cref{fig:super}), so a fast coarse scan can already capture a faithful field. The reconstructed field (or its $S(x,y)$) can then direct a second, finer pass to locations where diffraction actually varies: boundaries, triple junctions, and heterogeneous regions of interest, e.g., those that concentrate micromechanical damage \cite{stinville2017microstructural,latypov2021insight}. This approach can save acquisition time by avoiding full-density scans of featureless grain interiors. However, one should remember that off-grid reconstruction cannot create new information. Features smaller than the coarse measurement step (e.g., fine twins, precipitates, grain nuclei) cannot be recovered by neural representation even if it infers patterns at any resolution. 

Efficient full-pattern preservation is the second enabled opportunity. The trained model  reconstructs either dataset at the individual pattern level to the fidelity shown in \Cref{sec:recon}. So SIREN weights plus \num{512} learned basis patterns taking \SI{34.8}{\mega\byte} on disk replace \SI{26.0}{\giga\byte} of raw patterns of each EBSD dataset. For comparison, a generaic and lossless image compression of Kikuchi patterns achieves factors of only a few; storing per-point PCA coefficients at the same rank (with no network) takes about \SI{1.9}{\giga\byte} per scan ($\sim$50 times the TF-INR model) and remains grid-bound. The reduction with TF-INR is lossy in a controlled sense: its error budget is exactly the reconstruction error reported above (\SIrange{0.6}{0.8}{\percent} MAE of full intensity). A microscopy operator can target a desired level of fidelity at the time of training an INR. Compression with controlled fidelity is of significance because advanced EBSD workflows require full Kikuchi patterns (dictionary indexing,  autoencoder methods, see \Cref{sec:intro}). Yet, in current practice, the patterns are routinely discarded after indexing for orientations, especially in industrial settings where the high measurement throughput makes storage of raw patterns prohibitively expensive. Nearly three orders of magnitude reduction demonstrated with TF-INR could resolve the storage bottleneck and make patterns available on demand from saved model weights. It also eases data sharing and archiving for reproducibility: light-weight trained model can be more readily uploaded to repositories, from which collaborators could reconstruct the patterns on demand and run any downstream analysis on full patterns. For these reasons, we deliberately trained on raw patterns (even without background removal) so that users have a freedom to apply any pattern processing of their choosing upon reconstruction (including methods not yet developed).

Measured inference time is favorable for on-demand reconstruction: it takes only \SI{1.5}{\second} of GPU or \SI{11.6}{\second} of CPU time to reconstruct a full scan of $9\times10^5$ patterns from the TF-INR models trained here (about \SI{0.6}{\milli\second} per pattern). An interactive web application further demonstrates nearly instantaneous and spatially continuous query of the learned diffraction field (link below). 

\subsection{Limitations}

\noindent \textit{Training from scratch for each EBSD scan} The present approach trains an individual INR model for each EBSD scan. For the large datasets considered here (\num{9e5} patterns) and using $R=512$, it takes about three hours to train a TF-INR instance on a workstation with NVIDIA RTX A4000. While it is a one-time investment for on-demand pattern-level reconstruction, the training is not as instantaneous as indexing. Possible mitigations include (i) fixing the bases to a simulated pattern dictionary \cite{chen2015dictionary}, which removes basis optimization entirely and ties the coefficients to known orientations; (ii) pre-training coefficient networks on maps of the same or a similar material so that a new dataset requires fine-tuning instead of full training.

\noindent \textit{The rank is a microstructure-dependent hyperparameter} We used a fixed rank $R=512$ for both microstructures, although PCA with 512 components captures different levels of variance (\Cref{fig:methods}a). Strongly deformed, ultrafine-grained, or multiphase materials may need a higher (and individually determined) rank for faithful representation, compared to the single-phase f.c.c.\ alloy considered here. Automating the choice against target fidelity is a direct extension of the present PCA-guided procedure. Optimal rank determination should weigh the reconstruction accuracy against intentional under-ranking as an EBSD denoiser as our tests suggest that lower rank representation can discard artifacts. 

\noindent \textit{Off-grid patterns are inferences} The super-resolution capability is validated against held-out measured patterns (\Cref{fig:super}); the sub-grid pattern evolution across boundaries (\Cref{fig:interp}) has no ground truth in this study. Its character is physically realistic with the gradual exchange of band systems matching diffraction from two lattices sharing the interaction volume \cite{tong2015effect,marquardt2017quantitative}. However, this sub-grid behavior needs confirmation by measurement at finer scale, e.g., by transmission Kikuchi diffraction or high-resolution EBSD \cite{trimby2012orientation,keller2012transmission,wilkinson2006high}.

\noindent \textit{The diffraction gradient combines multiple effects} $S(x,y)$ incorporates orientation change, band-sharpness change, and contrast change into one magnitude. Once combined into a single scalar $S$, the contributions of individual effects cannot be recovered. Also, the magnitude is relative and meaningful only within one trained representation (but not across maps). These limitations can be addressed. Derivatives of individual coefficients on physics-based bases (e.g., from dictionaries) would maintain crystallographic meaning. Alternatively, a hybrid workflow could be devised where $S$ localizes positions of strong discontinuity with consequent conventional indexing or additional fine-grid measurement discussed in \Cref{sec:disco-2}.

\subsection{Outlook}

Neural EBSD demonstrated here represents an instance of a broader paradigm of \textit{differentiable characterization}: the replacement of a grid-based measurement by a continuous, differentiable neural surrogate of a sampled material field. The presented framework and methods are not limited to EBSD and can apply wherever a scan records an image or a spectrum at every specimen position. It can thus readily extend to  transmission Kikuchi diffraction, 4D-STEM, EELS, EDS, and other advanced, spatially resolved characterization methods. For EBSD, immediate extensions include (i) 3D EBSD \cite{echlin2015tribeam,zaefferer2008three}, for which the coefficient field is generalized to three dimensions, $A(x,y,z)$, while the 3D gradient $S(x,y,z)$  localizes boundary surfaces; (ii) multiphase materials, which can test the rank capacity;  (iii) relating bases to simulations or dictionaries (which will give the coefficient derivatives crystallographic meaning); and (iv) INR-in-the-loop acquisition, in which a field trained to a coarse scan directs subsequent measurement.

\section{Conclusions}

We introduced neural EBSD, which treats a spatially resolved EBSD measurement as a continuous, differentiable field of Kikuchi diffraction intensity instead of a set of independent patterns fixed to the acquisition grid.  We developed two formulations of the field: a joint network over all four coordinates (4D-INR) and a tensor-factorized representation combining continuous specimen-domain coefficient fields with learned detector-domain bases (TF-INR). Both formulations were evaluated on recrystallized (RX718) and additively manufactured (AM718) Inconel 718, which resulted in the following findings:

\begin{enumerate}

\item \textit{Factorization makes the field learnable.} TF-INR represents both microstructures with map-averaged errors below \SI{1}{\percent} of full-scale intensity, reproducing indexed orientations to within \SI{4}{\degree} at \qty{98.8}{\percent} of \num{9e5} scan positions in RX718, and outperforms the joint formulation in the present implementations. The learned bases are not abstract components: many index as valid Kikuchi patterns, and the coefficient fields recover the grain structure without being given orientation or boundary information.

\item \textit{Boundaries and defect substructure localize without indexing.} Analytical differentiation of the coefficient fields yields a diffraction gradient field $S(x,y)$ that is continuous in the specimen frame and free of indexing, disorientation thresholds, and staircase artifacts. It recovers the boundary network of RX718 and, in AM718, resolves high- and low-angle boundaries together with intragranular heterogeneity consistent with the dislocation-cell substructure of the as-built condition. It is a single scalar field computed from raw patterns that exposes features that conventional workflows obtain (if at all) from separate analyses and algorithms.

\item \textit{The field can be queried between measurements.} Trained on one quarter of the scan positions, TF-INR recovered withheld patterns whose indexed orientations agreed with reference to within \SI{4}{\degree} at \SI{97}{\percent} of positions in RX718. Querying four positions between adjacent scan points refines the spatial resolution five-fold along a chosen path and reproduces the gradual exchange of band structures expected across a grain boundary; $S(x,y)$ and the confidence metric place the boundary at the same sub-grid position.

\item \textit{The representation is a compact surrogate for the data.} Storing the network weights and learned bases in place of the raw patterns reduces each dataset from about \SI{26.0}{\giga\byte} to \SI{34.8}{\mega\byte} ($\sim$\num{745}-fold reduction, lossy within the reconstruction error) while keeping every pattern reconstructable in milliseconds on demand for downstream full-pattern analyses.

\end{enumerate}

More broadly, this work shows that the grid imposed by the experiment need not be inherited into the representation and analysis of fundamentally continuous microstructure fields. This work therefore serves as a demonstration of a more general paradigm of differentiable characterization transferable to other spatially resolved characterization methods and modalities. 

\section*{Data and code availability}

% The code implementing 4D-INR and TF-INR and the trained models for both datasets are available at \url{https://github.com/...}. 
An interactive web demonstration of continuous query of the learned diffraction field is available at \url{https://neural-ebsd.github.io/}. The EBSD datasets analyzed in this work were published by Calvat et al.\ \cite{calvat2025kikuchipattern}.

\section*{Acknowledgments}

% MIL acknowledges the support by the National Science Foundation under Award No.\ 2441813. 
The authors thank Drs.\ J.C.\ Stinville and Mathieu Calvat at UIUC and Drs.\ Pascal Thome and I-Ting Ho at the University of Arizona for insightful discussions.

% ==============================================================================
% TRANSITION TO SUPPLEMENTARY MATERIAL
% ==============================================================================

\clearpage % Start on a new page

% 1. Reset figure counter to 0
\setcounter{figure}{0}

% 2. Update figure counter representation for captions (\begin{figure}...)
\renewcommand{\thefigure}{S\arabic{figure}}

% 3. Update cleveref label format for references (\Cref{...})
\crefname{figure}{Figure}{Figures} 
% Note: Because \thefigure now outputs "S1", \Cref{fig:supp_example} will produce "Figure S1"

% (Optional) Reset and prepend table numbering as well if needed:
% \setcounter{table}{0}
% \renewcommand{\thetable}{S\arabic{table}}
% ==============================================================================
\begin{figure*}[h!]
  \centering
  \includegraphics[width=\textwidth]{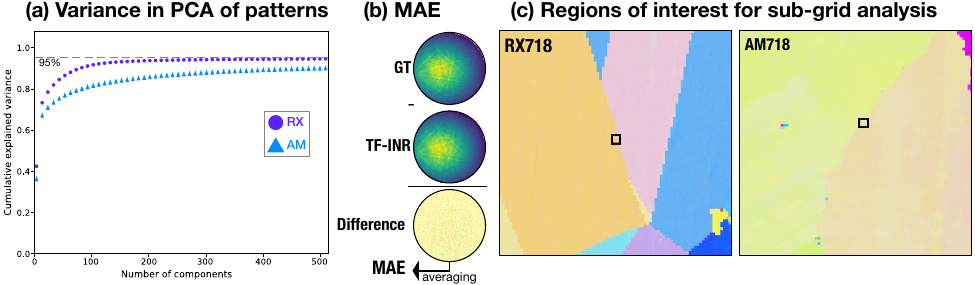}
  \caption{ (a) Explained variance of randomly sampled \num{200000} patterns from the two EBSD datasets; (b) visual illustration of pattern-level MAE; (c) regions of interest showing the location of $3\times3$ windows for sub-grid sampling in \Cref{fig:interp}.}
  \label{fig:methods}
\end{figure*}

\section*{Supplementary materials}

\subsection*{Evaluation metrics}

We evaluate the reconstruction quality of the INR models using pattern level intensity error and orientation consistency. 

\paragraph{Mean absolute error (MAE)}

Let  $I(x,y,u,v)$ denote the reference experimental Kikuchi pattern intensity, and $\hat{I}(x,y,u,v)$ denote the corresponding INR-reconstructed intensity. Detector pixels are evaluated within the circular aperture mask $\mathcal{M}$. 

The mean absolute error (MAE) per pattern (\Cref{fig:methods}b) is defined as

\begin{multline}
\label{eq:mae_pattern}
    \mathrm{MAE}(x,y)= \\
    \frac{1}{|\mathcal{M}|}
    \sum_{(u,v)\in\mathcal{M}}
    \left| \hat{I}(x,y,u,v) - I(x,y,u,v) \right|.
\end{multline}

The map-averaged MAE is then computed by averaging over all scan positions,

\begin{equation}
    \overline{\mathrm{MAE}} = \frac{1}{N} \sum_{(x,y)}
    \mathrm{MAE}(x,y),
    \label{eq:mae_map}
\end{equation}

\noindent where $N$ is the total number of scan points.

\paragraph{Structural similarity index measure (SSIM)}

\begin{figure*}[h!]
  \centering
  \includegraphics[width=\textwidth]{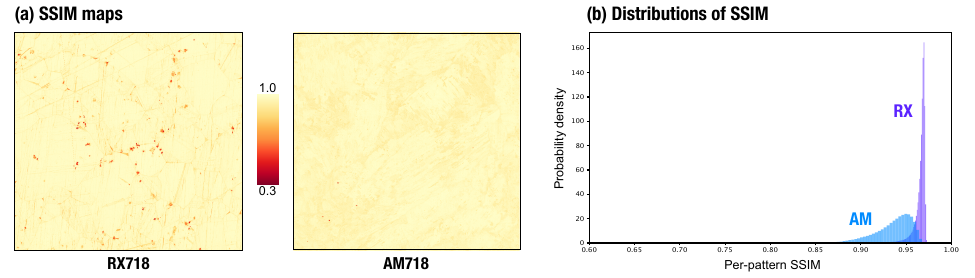}
  \caption{SSIM analysis of patterns reconstructed by TF-INR for RX718 and AM718 visualized as (a) maps and (b) distributions.}
  \label{fig:ssim}
\end{figure*}

SSIM \cite{wang2004image} compares reconstructed and measured patterns through local statistics computed using an $11\times11$ Gaussian-weighted sliding window with $\sigma=1.5$ pixels within the aperture mask $\mathcal{M}$. For a pair of co-located windows $p$ (measured) and $\hat{p}$ (reconstructed),

% \begin{equation}
%     \mathrm{SSIM}(p,\hat{p}) =
%     \frac{\left(2\mu_p\mu_{\hat{p}} + C_1\right)
%           \left(2\sigma_{p\hat{p}} + C_2\right)}
%          {\left(\mu_p^2 + \mu_{\hat{p}}^2 + C_1\right)
%           \left(\sigma_p^2 + \sigma_{\hat{p}}^2 + C_2\right)},
%     \label{eq:ssim}
% \end{equation}

\begin{multline}
\label{eq:ssim}
    \mathrm{SSIM}(p,\hat{p}) = \\
    \frac{\left(2\mu_p\mu_{\hat{p}} + C_1\right)
          \left(2\sigma_{p\hat{p}} + C_2\right)}
         {\left(\mu_p^2 + \mu_{\hat{p}}^2 + C_1\right)
          \left(\sigma_p^2 + \sigma_{\hat{p}}^2 + C_2\right)},
\end{multline}

\noindent where $\mu$ and $\sigma^2$ are the local mean and variance, $\sigma_{p\hat{p}}$ is the local covariance, and $C_1=(K_1L)^2$, $C_2=(K_2L)^2$ are stabilizing constants with $K_1=0.01$, $K_2=0.03$, and $L$ is the scan-specific intensity range of the 16-bit patterns. For the intensity range in RX718 ($[7674, 61183]$): $L=53509$; for AM718 with intensity range  $[1681, 58727]$: $L=57046$. The pattern-level value $\mathrm{SSIM}(x,y)$ is the mean of \Cref{eq:ssim} over all windows within $\mathcal{M}$, and the map-averaged $\overline{\mathrm{SSIM}}$ follows \Cref{eq:mae_map}. Values approach unity for identical patterns. \Cref{fig:ssim} shows maps and distributions of SSIM for patterns reconstructed with TF-INR for the RX718 and AM718 microstructures.

\paragraph{Orientation deviation}

For Hough indexing, both experimental and reconstructed Kikuchi patterns were Gaussian bandpass filtered to reduce background, using $\sigma_\mathrm{low}=0.25$ and $\sigma_\mathrm{high}=0.035$ in Nyquist frequency units \cite{thome2019ni}.

For each scan point $(x,y)$, two orientations are obtained: one, $g_\text{ref}$, by Hough indexing of the measured Kikuchi pattern (reference) and one, $g$, by Hough indexing of the reconstructed pattern from neural representation. The orientation deviation, $\Delta\theta$, is computed in MTEX \cite{bachmann2010texture} as:

\begin{equation}
    \Delta\theta(x,y) = \angle\!\left(
    g_{\mathrm{ref}}^{-1}g
    \right),
    \label{eq:disorientation}
\end{equation}

\noindent where $\angle(\cdot)$ denotes the cubic symmetry-reduced rotation angle.

\begin{figure*}[h!]
  \centering
  \includegraphics[width=\textwidth]{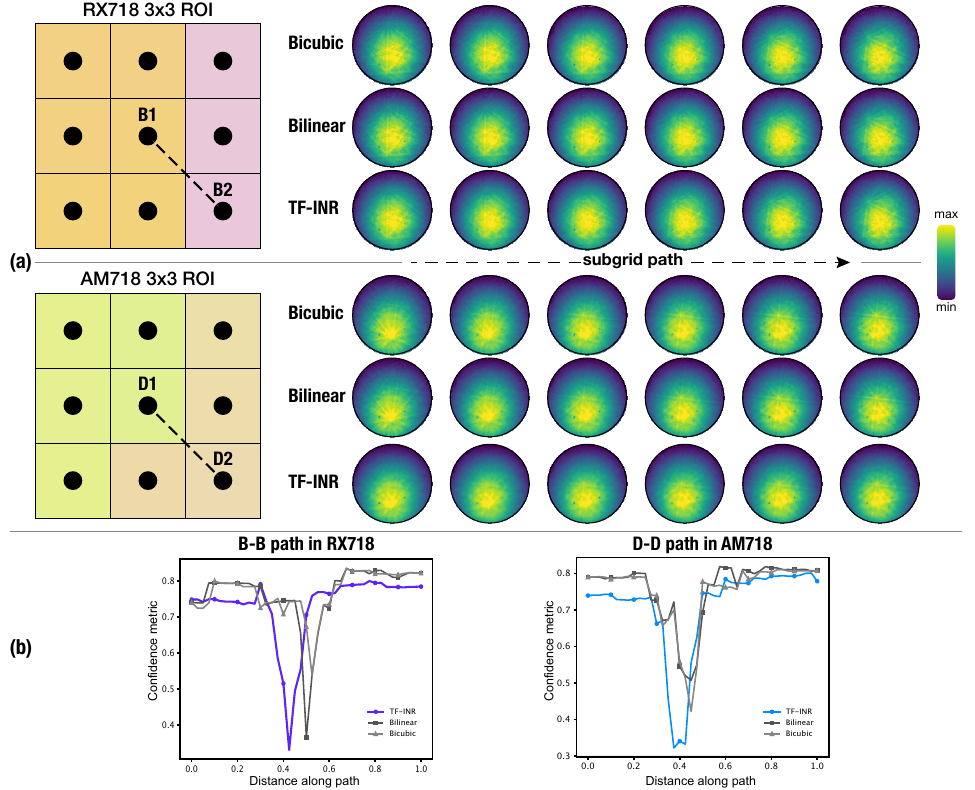}
  \caption{Comparison of TF-INR query with traditional interpolations for the diagonal sub-grid paths analyzed in \Cref{fig:interp}: (a) off-grid patterns and their paths; (b) indexed confidence metric for these patterns. Note the dips off the midpoints (distance 0.5) for TF-INR patterns along both paths. }
  \label{fig:supp_interp-comp}
\end{figure*}

\paragraph{Confidence metric} Hough indexing also returns a confidence metric for each pattern, which measures how strongly the best orientation solution is favored over competing solutions in the band-voting step (normalized to [0,1]). Low values indicate ambiguous patterns, such as those formed by overlapping diffraction from two grains near a boundary \cite{wright1992automatic, anes2026kikuchipy}. We compute the confidence metric  in \texttt{kikuchipy} and use in \Cref{sec:interp,sec:gbs} to track diffraction ambiguity along sub-grid paths.

\subsection*{Band region identification}

Band regions were identified using up to \num{128} representative patterns sampled across the scan area. For each normalized pattern $I_k(u,v)$, the detector-plane gradient magnitude within the aperture mask $\mathcal{M}$ was computed as

\begin{equation}
G_k(u,v)
=
\sqrt{
\left(\frac{\partial I_k}{\partial u}\right)^2
+
\left(\frac{\partial I_k}{\partial v}\right)^2
}.
\label{eq:band_gradient}
\end{equation}

For each pattern, pixels within the top 15\% of $G_k$ were marked in a binary map. These maps were averaged to measure how consistently each pixel exhibited a strong gradient across the representative patterns. The 15\% of pixels with the highest occurrence were defined as the pixels most likely to correspond to band regions. These pixels were assigned $w=2$ during training, while all remaining pixels were assigned $w=1$.

\begin{figure*}[h!]
  \centering
  \includegraphics[width=\textwidth]{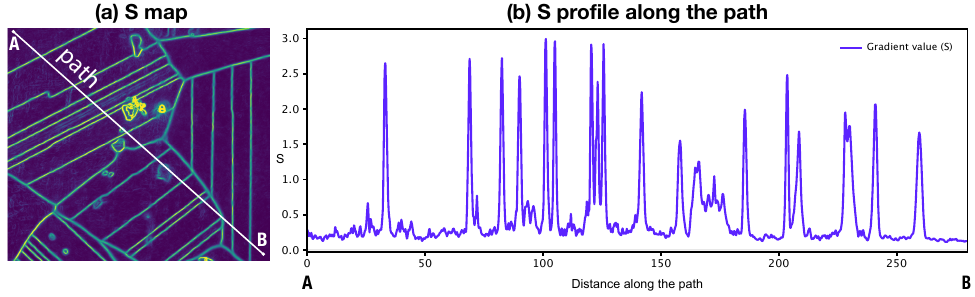}
  \caption{Diffraction gradient, $S(x,y)$ shown as (a) map for a representative region in the RX718 microstructure; (b) profile along the path shown in (a). Note that boundaries correspond to clear peaks in $S$ suggesting segmentation feasibility.}
  \label{fig:supp_gradient}
\end{figure*}

\bibliography{refs}
\end{document}